\documentclass[traditabstract]{aa}

\usepackage{graphicx}
\usepackage[varg]{txfonts}
\usepackage{natbib,twoopt}
\usepackage{lscape}
\usepackage{subcaption}
\usepackage{nicefrac}
\bibpunct{(}{)}{;}{a}{}{,} 

\usepackage[draft]{hyperref}

\makeatother\bibpunct[, ]{(}{)}{;}{a}{}{,}

\newcommand{\farc}{\hbox{$.\!\!^{\prime\prime}$}} 
\newcommand{\farcmin}{\hbox{$.\!\!^{\prime}$}}

\newcommand{\hb}{H$\beta$} 
\newcommand{\ha}{H$\alpha$}
\newcommand{\hg}{H$\gamma$} 
\newcommand{\hd}{H$\delta$} 
 
\newcommand{\hii}{\mbox{H~{\sc ii}}} 
 
\newcommand{\oh}{12+\log(\mathrm{O/H})}

\newcommand{\hei}{\ion{He}{i}} 
 
\newcommand{\sii}{[\ion{S}{ii}]}

\newcommand{\oiii}{[\ion{O}{iii}]}

\newcommand{\nii}{[\ion{N}{ii}]}

\begin{document}

\title{The environment of the SN-less GRB~111005A at $z=0.0133$\thanks{Based on observations collected at the ESO Paranal observatory under ESO program 60.A-9330(A).}}
\titlerunning{The SN-less host galaxy of GRB~111005A}

\author{M. Tanga\inst{1}
\and T.~Kr\"{u}hler\inst{1}
\and P.~Schady \inst{1} 
\and S.~Klose \inst{2}
\and J. F. Graham\inst{1,3}
\and J. Greiner\inst{1}
\and D. A. Kann\inst{4}
\and M. Nardini\inst{5}}
\institute{Max-Planck-Institut f\"{u}r extraterrestrische Physik, Giessenbachstra\ss e, 85748 Garching, Germany \and 
Th\"uringer Landessternwarte Tautenburg, Sternwarte 5, 07778 Tautenburg, Germany \and
Kavli Institute for Astronomy and Astrophysics, Peking University, Beijing 100871, PR China \and
Instituto de Astrof\'{i}sica de Andaluc\'{i}a – CSIC, Glorieta de la Astronom\'{i}a s/n, 18008 Granada, Spain \and
Universit\`{a} degli studi di Milano-Bicocca, Piazza della Scienza 3, 20126, Milano, Italy
}

\abstract{The {\em collapsar} model has proved highly successful in explaining the properties of long $\gamma$-ray bursts (GRBs), with the most direct confirmation being the detection of a supernova (SN) coincident with the majority of nearby long GRBs. Within this model, a long GRB is produced by the core-collapse of a metal-poor, rapidly rotating, massive star. The detection of some long GRBs in metal-rich environments, and more fundamentally the three examples of long GRBs (GRB~060505, GRB~060614 and GRB~111005A) with no coincident SN detection down to very deep limits is in strong contention with theoretical expectations. In this paper we present MUSE observations of the host galaxy of GRB~111005A, which is the most recent and compelling example yet of a SN-less, long GRB. At $z=0.01326$, GRB~111005A is the third closest GRB ever detected, and second closest long duration GRB, enabling the nearby environment to be studied at a resolution of 270~pc at $z=0.01326$. From the analysis of the MUSE data cube, we find GRB~111005A to have occurred within a metal-rich environment with little signs of ongoing star formation. Spectral analysis at the position of the GRB indicates the presence of an old stellar population ($\tau \ge$ 10 Myr), which limits the mass of the GRB progenitor to M$_{ZAMS}<15$~M$_{\odot}$, in direct conflict with the collapsar model. Our deep limits on the presence of any SN emission combined with the environmental conditions at the position of GRB~111005A necessitate the exploration of a novel long GRB formation mechanism that is unrelated to massive stars.}

\maketitle

\section{Introduction}
\label{sec:Intro}

The connection between long-duration, soft-spectrum $\gamma$-ray bursts (GRBs), \cite[see][for a review]{2009ARA&A..47..567G} and core-collapse supernovae (SNe) is arguably the most fundamental clue to the nature of GRB progenitors and firmly links GRBs to the death of massive stars \citep[e.g.][]{2003Natur.423..847H,2003ApJ...591L..17S}. However, the exact properties of the progenitor star and the explosion mechanism remain unknown, and both metallicity and high angular momentum likely play a critical role in the formation of a GRB. 

The popular collapsar model for long GRBs, for example, requires their progenitors to be rapidly rotating massive stars \citep[zero-age main sequence mass $M_{\mathrm{ZAMS}} \sim 30~M_{\odot}$;][]{1993ApJ...405..273W, 1999ApJ...524..262M}. However, massive stars lose mass, and spin down due to their strong winds. The mass-loss rate is metallicity dependent ($\dot{M}\propto Z^{0.86}$) ~\citep{2005A&A...442..587V}, and thus in order to maintain the massively rotating cores required to form a GRB, the simplest collapsar model requires the GRB progenitor to be metal poor $(Z < 0.3 - 0.5~Z_\odot)$ \citep{2003ApJ...591..288H,2005A&A...443..581H,Woosley2006ApJ}. Alternatively GRBs could form due to unusual evolution in binary systems. A more massive star in a binary system could either be tidally spun up \citep[e.g.][]{Detemers2008AA} or helium cores in binary systems could merge to form GRBs \citep[e.g.][]{Izzard2004MNRAS, Podsiadlowski2004ApJ, Podsiadlowski2010MNRAS}.

The properties of the GRB progenitor stars, in particular high-mass and low metallicity, should also be reflected in their immediate environment, and therefore, GRB hosts have been the target of extensive observations \citep[e.g.][]{2003A&A...400..499L, 2004A&A...425..913C, 2006Natur.441..463F, 2009ApJ...691..182S}. However, it was only with the advent of the The Neil Gehrels \textit{Swift} Observatory \citep{2004ApJ...611.1005G}, and the subsequently significantly improved GRB rapid localizations provided by the onboard X-ray telescope \cite[XRT;][]{2005SSRv..120..165B} and the Ultraviolet and optical telescope \cite[UVOT;][]{2005SSRv..120...95R} that GRB afterglow and host samples were able to reach statistically significant numbers, and previous sample biases were able to be addressed \citep{2012ApJ...756..187H, 2016ApJ...817....7P}. Recent sample studies show indeed that the ratio between GRB formation and star-formation falls dramatically above a certain metallicity, with the exact value $(0.5 - 1.0 Z_\odot)$ being somewhat debated in the literature \citep[e.g.][]{2013ApJ...774..119G, 2015A&A...581A.125K, 2016ApJ...817....7P, 2017A&A...599A.120V, 2017ApJ...834..170G}. However, a non-negligible fraction of long GRB hosts with a metallicity around the solar value \citep[e.g.][]{2009ApJ...691L..27P, 2012MNRAS.420..627S, 2013A&A...556A..23E, 2015A&A...579A.126S} does exist, in apparent conflict with the collapsar model.

With the exception of space-based imaging \citep{2006Natur.441..463F, 2016ApJ...817..144B, 2017MNRAS.467.1795L}, the bulk of information on GRB environments has been thus far obtained by integrating the physical properties over the full galaxy. These data do not probe the actual properties of the \hii-region where the progenitor star was born, and only sensitive imaging data and integral-field spectroscopy with a high spatial resolution are capable of probing the small length scales within GRB hosts necessary to resolve individual star-forming regions. There are now a number of GRB host galaxies that have been observed with integral-field spectrographs \citep[e.g.][]{2008A&A...490...45C,2014MNRAS.441.2034T,2017MNRAS.472.4480I, 2017A&A...602A..85K}. However, for the most part the observations have only resolved the host galaxy on kpc scales, where the emission from several \hii\ regions, containing numerous stellar populations, are blended. The exception to this was the recent MUSE observations of the host galaxy of GRB 980425/SN1998bw \citep{2017A&A...602A..85K}, which, at $z=0.0085$, is the nearest GRB ever detected. The MUSE data resolve the host galaxy down to 100~pc, and even though we do not observe the metallicity of the GRB progenitor directly, with such data we can isolate individual \hii-regions within the host, and estimate the metallicity of the region. Within this region one can expect the gas to be well mixed, allowing us to obtain robust constraints on the stellar population age and \hii-region oxygen abundance, and thus progenitor mass and metallicity \citep{2017A&A...602A..85K}

A further twist in the search for GRB progenitors appeared in 2006 with the surprising discovery of two apparently long GRBs (GRB~060505 and GRB~060614) at low redshift ($z=0.089$ and $z=0.125$, respectively) where a luminous supernova akin to the GRB-SN poster-children of GRB~980425/SN~1998bw \citep{1998Natur.395..670G}, GRB~030329/SN~2003dh \citep[e.g.][]{2003Natur.423..847H} or GRB~060218/SN~2006aj \citep[e.g.][]{2006Natur.442.1011P}, would have been discovered if it were present \citep{2006Natur.444.1047, 2006Natur.444.1053G, 2006Natur.444.1044G, 2006Natur.444.1050D,2007ApJ...662.1129O}. GRB~060505 and GRB~060614 were suggestive of a new type of explosion process for some long GRBs, consisting either of the direct collapse of a massive star to a black hole or a long-lived central engine that is unrelated to a massive star. However, simpler alternatives are not strictly ruled out: there is always the possibility of a wrong redshift\footnote{The redshifts for both GRBs stem from host identifications, which are sometimes prone to errors \citep[e.g.][for such a case]{2017MNRAS.465L..89P}.} and it has also been argued that GRB~060614 shares more properties with the short rather than the long class of bursts \citep[e.g][]{2007ApJ...655L..25Z,2007ApJ...662.1129O,2011ApJ...734...96K} \citep[but see][]{2008ApJ...677L..85M, 2008ApJ...676.1151T}.

In this paper we analyse the third, and nearest SN-less long GRB, GRB~111005A at $z=0.0133$. We discuss the prompt $\gamma$-ray emission as observed by \textit{Swift}/BAT, provide deep limits on the supernova from the Gamma-Ray Optical/Near-infrared Detector (GROND, \citealt{2008PASP..120..405G}), and use high spatial-resolution integral-field spectroscopy with the Multi-Unit Spectroscopic Explorer (MUSE, \citealt{2010SPIE.7735E..08B}) to study the GRB environment in exquisite detail. Due to its proximity, GRB~111005A provides an important window to study the progenitors of GRBs, in particular those of SN-less long GRBs. In Sections~\ref{sec:grb} and~\ref{sec:env}~we provide details on GRB~111005A, its galactic environment as well as our observations and data analysis respectively. We discuss and summarize our results on the environmental and implied progenitor properties of GRB~111005A in Section~\ref{sec:dis}.

Throughout this article, we assume concordance $\Lambda$CDM cosmology with the following parameters ($H_0=67.3\,\mathrm{km}\,\mathrm{s}^{-1}\,\mathrm{Mpc}^{-1}$, $\Omega_\mathrm{m}$=0.315, $\Omega_\Lambda$=0.685, \citealt{2014A&A...571A..16P}), a \citet{2003PASP..115..763C} initial mass function (IMF), and error bars represent the 1~$\sigma$ confidence interval.

\section{The gamma-ray burst~111005A}
\label{sec:grb}

\subsection{Prompt emission and initial follow up}

\begin{figure}
\includegraphics[angle=0, width=0.99\columnwidth]{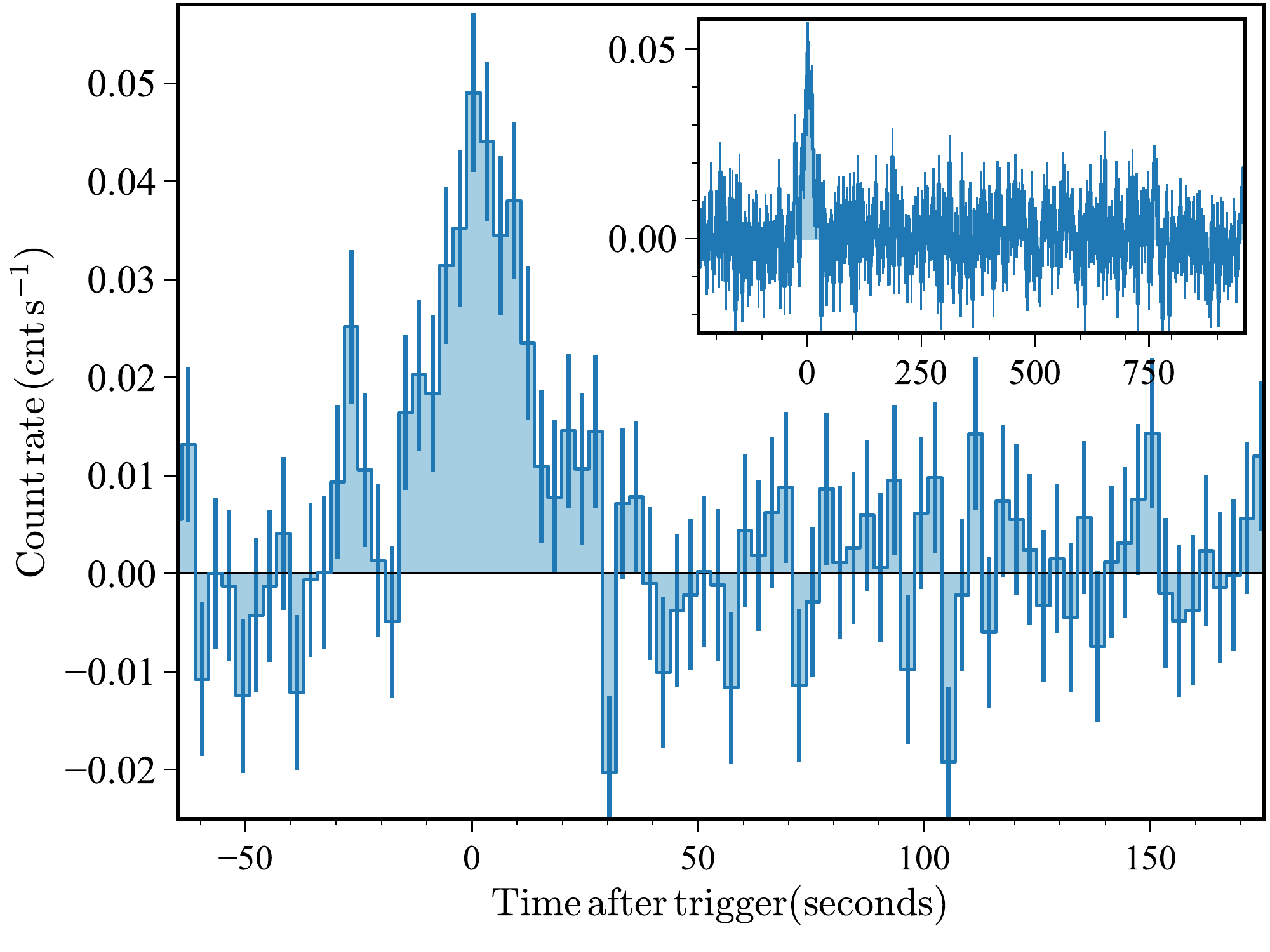}
\caption{\textit{Swift}/BAT lightcurve or GRB~111005A characterized by a single main peak and a $T_{90}=27\pm 8$~s.}
\label{fig:lc}
\end{figure}

{\em Swift} triggered on GRB 111005A at 08:05:14 UT on 5th October 2011 \citep{2011GCN..12413...1S}. However, due to a Sun observing constraint \textit{Swift}  was not able to slew to the GRB before the 25th December 2011, and no UVOT or XRT observations were thus taken. The prompt emission was relatively weak and it was predominantly detected in the 15-50 keV BAT energy range. The 15-50 keV mask-weighted light curve consists of a single main peak starting at roughly $T-$15 sec (Fig.~\ref{fig:lc}), with a $T_{90} = 27\pm 8$~s (estimated error including systematics), and a fluence over this time of $3.1\times 10^{-7}$~erg~cm$^{-2}$. The spectrum taken over $T_{90}$ is best-fit by a power-law index with $\Gamma = 2.4\pm0.2$ ($\chi^2/\mathrm{dof}=44.32/56$) and the hardness ratio is $F_{\mathrm{(50-150~keV)}}/F_{\mathrm{(15-50~keV)}} = 0.56$. These temporal and spectral properties put GRB~111005A firmly into the class of long-soft GRBs. It is worth noting that that applying the cross-correlation function between the 15-25 keV and 25-50 keV mask-weighted light curves implies a negligible time-lag on the order of 0.01~sec. However, whereas short GRBs have almost exclusively negligible time-lags, long GRBs can have lags ranging from zero to several seconds \citep{2002ApJ...579..386N,2006ApJ...643..266N,2007ApJS..169...62H}. The negligible time-lag measured in the prompt emission of GRB~111005A does not, therefore, help discriminate between the classification of this burst in terms of short and long GRBs.

Under the assumption that $\Gamma$ is the high-energy power-law index of the GRB's synchrotron spectrum (and the peak of the spectrum is in the lower energy band of the BAT energy response), the isotropic energy of GRB~111005A is $E_{\mathrm{iso}} \sim 5 \times 10^{47}$~erg with errors of around 50\% mostly because of the uncertainty in the high-energy spectrum. Despite these uncertainties, GRB~111005A is clearly a low-luminosity GRB (if the GRB association with ESO 580-49 is correct, see Section~\ref{sec:assoc}), although, it is more luminous that the archetypal GRB~980425 by around an order of magnitude. Irrespective of the relative luminosity of GRB~111005A, current observations of SNe associated with low-luminosity and cosmological GRBs show a comparatively small range in SN peak magnitudes, spanning $2-3$ magnitudes, in contrast to the $\sim 6$ magnitudes in isotropic luminosity covered by GRBs \citep[e.g.][]{2012PASJ...64..115N,2013RSPTA.37120275H,2014A&A...567A..29M,2014ApJ...781...37P}. This therefore suggests that the low luminosity nature of GRB~111005A does not necessarily imply a comparatively low luminosity accompanying SN, especially given that it was in fact more luminous than GRB~980425.

The 90\% confidence \textit{Swift}/BAT error-circle of 2\farcmin{1} radius \citep{2011GCN..12415...1B} contained ESO 580-49, a bright spiral galaxy seen nearly edge on, which if the host galaxy, would make GRB~111005A the second closest GRB detected with \textit{Swift}, and the third closest GRB ever detected (after GRB~980425 and GRB~170817A).

Early optical and near-infrared observations \citep[][and Section~\ref{sec:snlim}]{2016arXiv161006928M} did not find a variable source in the BAT error-circle, but a relatively bright radio transient was identified a couple of days later, close to the centre of ESO 580-49 \citep{2016arXiv161006928M}. The magnitude of the radio emission of the transient in the frequency range 9-18 GHz was 10-100 times fainter than that observed in other low-$z$ GRBs such as GRB~031203, GRB~060218 and GRB~980425, it further showed a plateau phase followed by sharp decay with the break occurring after a month rather than a few days. However, a lack of flaring and the similarity of the afterglow with that of GRBs gives credibility to the association of the radio-transient with the GRB. 

\subsection{The association with ESO 580-49}
\label{sec:assoc}

All considerations that follow in this paper fundamentally rely on the fact that GRB~111005A indeed exploded in ESO 580-49. This requires two different assumption to be valid: that the GRB prompt emission is associated with the radio transient, and that the radio transient is associated with ESO 580-49.

To estimate the statistical probabilities for each of these associations we follow \citet{2017MNRAS.465L..89P}, using similarly the rate of radio transients from \citet{2011ApJ...740...65O}. Assuming a typical timescale of the radio transient of 100~days, and peak brightness of 2~mJy \citep{2016arXiv161006928M}, we would expect around $\sim2\times 10^{-6}$ ($2\sigma$ upper limit of $10^{-5}$) radio transients in an area of the BAT error-circle within 5 days of the GRB. The sky density of similarly bright galaxies ($r\sim14$~mag) as ESO 580-49 is around 0.5~deg$^{-2}$, and assuming that each of them covers a sky area of $\sim$0.3 arcmin$^{2}$ again as ESO 580-49, leads to a probability of $3\times 10^{-5}$ that the position of a random point source falls onto the visible extent of a bright galaxy. Even taking into account that \textit{Swift} has detected and localized $>1000$\,GRBs the combined chance coincidence that any of these is wrongly associated with a bright galaxy by chance is thus only 0.02. 

While this number is considerably larger and likely more conservative than the one calculated by \citet{2016arXiv161006928M}, we agree that it is very reasonable to assume that the GRB is associated with the radio transient, and the radio transient with ESO 580-49. This places GRB~111005A at $z=0.0133$, which corresponds to a luminosity distance of $L_D=59$~Mpc and an angular scale of 0.29~kpc~arcsec$^{-1}$.

\subsection{Limits on supernova emission}
\label{sec:snlim}

\begin{figure}
\includegraphics[angle=0, width=0.99\columnwidth]{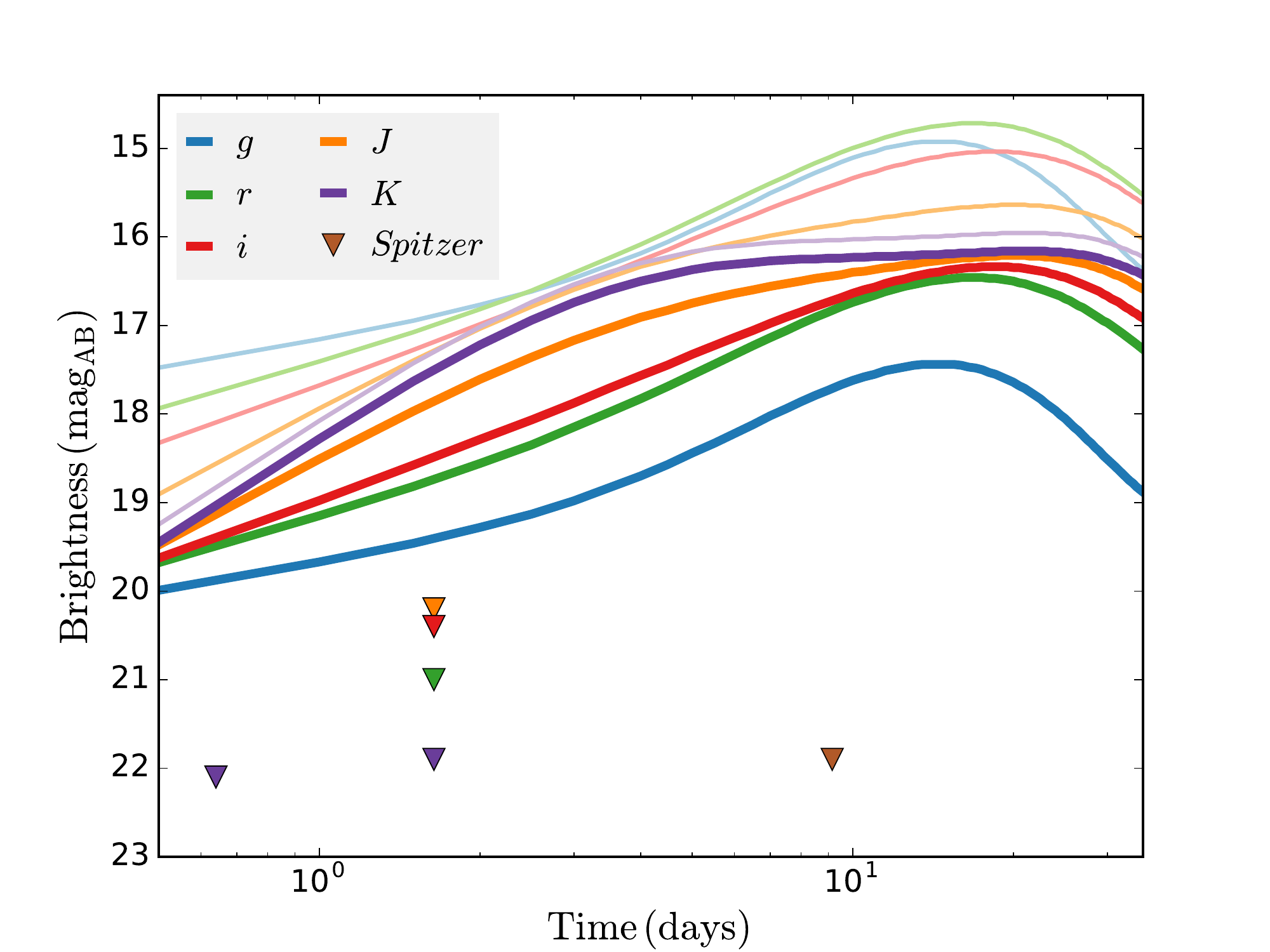}
\caption{$3\sigma$ upper limits on variable optical emission associated with GRB~111005A at the position of the radio transient. The solid, thick lines show the SN1998bw template as it would appear at the redshift of GRB~111005A and when reddened by a host $A_V=2$~mag assuming a Milky Way type reddening law with an $R_V=3.08$, while the thinner lines denote $A_V=0$~mag. The downward triangles show various upper limits on the SN emission, with the $riJ$ data from this work, and the $K$-band and \textit{Spitzer} upper limits from the data of \citet{2016arXiv161006928M}.}
\label{fig:sn}
\end{figure}

To derive limits on possible supernova emission following GRB~111005A, we use our own photometric follow-up as well as the data presented in \citet{2016arXiv161006928M}. Our observations were obtained during the nights of 2011-10-05 \citep{2011GCN..12417...1N} and 2011-10-06 with GROND, mounted on the 2.2~m MPG telescope in La Silla, Chile. Because the field of GRB~111005A was close to the Sun at the time of the GRB trigger, each epoch only consisted of a short sequence of dithered exposures in the $grizJH$ and $K_{\mathrm{s}}$ filters. These observations were complemented by images obtained several years after the event on 2017-04-26 that were 1-2 orders of magnitude deeper than the observations in 2011. We used these deeper images as a template for digital image subtraction. The data reduction and subtraction followed standard procedures \citep[details in e.g.][]{2008ApJ...685..376K}.

Constraining $3\sigma$ upper limits from GROND in the $riJ$ filters as well HAWK-I ($K$-band) and \textit{Spitzer} \citep{2016arXiv161006928M} are shown in Fig.~\ref{fig:sn}. To be able to compare them to the measurements of SN~1998bw at the same epoch, we use the light curve information compiled in \citet{2011AJ....141..163C}. The most direct comparison comes from the earliest two epochs (0.7~days, and 1.7 days after the GRB~980425) which were originally presented in \citet{1998Natur.395..670G}. These early observations show SN~1998bw at $V=15.9$ and with colours that indicate a relatively flat spectrum in $F_\nu$ ($B\sim16.0$~mag, $R_C\sim15.8$~mag). Unfortunately, no data in other photometric bands are available so early after the GRB~980425 trigger, so the comparison to other GROND filters comes from a temporal and spectral extrapolation with details given in \citet{2004ApJ...609..952Z}. 

The derived light-curves of SN1998bw shifted to $z=0.0133$ are shown in Fig.~\ref{fig:sn}: the solid thick lines describe the SN~1998bw light curve template for a host reddened by a Milky Way type extinction curve with $R_V=3.08$ and $A_{V}$ = 2 mag\footnote{We measure $A_{V} \sim$ 2 mag at the GRB position, see Section~\ref{sec:dustmap} for details.}, while thin lines describe the SN1998bw light curve template for an unreddened host ($A_{V}$ = 0 mag). An $A_{V} >$ 20 mag would be required for the \textit{Spitzer} data upper limit from \citet{2016arXiv161006928M} to lie above the SN light curve template. Thus our upper limits rule out a SN1998bw-like event for GRB~111005A. Any underlying SN must have been around 40 times fainter than SN1998bw in the $r$ band, corresponding to an absolute peak magnitude of $M_r>-12.8$. This is around 5 magnitudes fainter than typical GRB-SNe \citep{2013RSPTA.37120275H,2017AdAst2017E...5C}, and is around 0.6 mag deeper than the limits place on any SN associated with GRB~060614 \citep{2006Natur.444.1050D,2011ApJ...734...96K}. Only GRB~060505 has deeper limits on the presence of any simultaneous SN \citep[$M_R>-11$][]{2007ApJ...662.1129O}, although the classification of this GRB as short or long is unclear \citep[e.g.][]{2007ApJ...662.1129O,2011ApJ...734...96K}.

\section{The GRB environment}
\label{sec:env}
\subsection{Observations}

The field of the galaxy hosting GRB~111005A was observed using the Multi-Unit Spectroscopic Explorer (MUSE; \citealp{2010SPIE.7735E..08B}) during its science verification. MUSE is a state-of-the-art integral field unit (IFU) that continuously covers an area of 1 arcmin$^{2}$ in the optical wavelength range. It is mounted at the Unit Telescope 4 (Kueyen) of the ESO Very Large Telescope, and its combination of wavelength coverage, exquisite sensitivity, small spaxel size (0\farc{04}$^2$), and decent spectral resolving power $R$ ($R$ ranges between 1800 and 3600) is unprecedented among IFUs. In this way, MUSE delivers high-quality, spatially-resolved spectroscopy which allows us to study the environments of explosive transients at a high-level of detail.

Our observations were obtained on 2014-08-23, and consisted of four individual exposures of 630~s integration time each. The science frames were complemented by two pointings to blank sky for background subtraction, each of which had an integration time of 200~s. Sky conditions were clear, and we measure the image quality delivered by the telescope and instrument through the point-spread function (PSF) of stellar sources which have a full-width at half maximum (FWHM) of 1\farc{0} at 5000~\AA\ and 0\farc{85} at 9000~\AA.

The data were obtained with MUSE in its extended mode, which offers a somewhat bluer wavelength response (down to 4600\,\AA), with the compromise of second-order contamination in the wavelength regime redwards of 8000~\AA. The second-order contamination from the blue stellar continuum would cause an additional offset, while the second order of bright, blue emission lines such as \hd\ or \hg\ create a broad bump because the second order is unfocused in MUSE \citep{2015A&A...582A.114W}. Both of these effects are of no concern to us here, as we are mostly interested in narrow emission lines at wavelengths bluer than the \sii($\lambda\lambda$6717, 6731) doublet.

\subsection{MUSE data reduction}

The reduction of the integral-field spectroscopy follows closely what we have applied previously to MUSE observations \citep[e.g.][]{2016MNRAS.455.4087G, 2016arXiv160900013P, 2017A&A...602A..85K}. Very briefly, we used the ESO pipeline (\texttt{version 1.6.}, \citealp{2014ASPC..485..451W}) to correct for instrumental effects such as flat-fielding or inhomogeneous illumination within MUSE, and apply a wavelength, flux and astrometric calibration.

The initial sky-subtraction from the ESO pipeline was enhanced through the algorithms of \citet{2016MNRAS.458.3210S}. We have further used \texttt{molecfit} \citep{2015A&A...576A..77S} with a stellar spectrum in the MUSE field of view to derive a model of the telluric absorption and applied the resulting transmission correction to the full data cube. At this step, we also correct the spectra for the Galactic foreground extinction corresponding to a reddening $E_{B-V}=0.038$~mag, obtained from the re-calibrated \citet{1998ApJ...500..525S} maps \citep{2011ApJ...737..103S}.

Finally, we correct the MUSE astrometric solution, which shows a slight offset in both coordinates, through observations of the same field from the Hubble Space Telescope (HST). HST observed the field of GRB~111005A equipped with the Wide Field Camera (WFC3) in three different filters ($F438W$, $F606W$ and $F160W$). We download the respective images from the HST archive, and apply an astrometric solution by tying the centroid of field stars to their positions from the \textit{Gaia} catalog \citep{2016A&A...595A...1G, 2016A&A...595A...2G}. This procedure brings the WFC3 astrometry into the \textit{Gaia}-based system and has a root-mean-square (rms) scatter of less than 20 mas in both coordinates for all images. Finally, we use the WFC3-based positions of four field stars, which are also clearly detected in reconstructed MUSE images, to modify the VLT-based astrometry. A simple linear transformation in both coordinates yields an rms of 50 mas and brings the MUSE data into the \textit{Gaia} astrometric system, which in turn is aligned with the International Celestial Reference Frame (ICRF) with high precision \citep{2016A&A...595A...4L}. We can hence localize the precise, 0.2 mas accuracy VLBI radio position of the GRB~111005A afterglow \citep{2016arXiv161006928M}, within the MUSE data cube to a precision of better than one spaxel, where the MUSE linear pixel scale is 0\farc{2}.

In this astrometric reference frame, the centre of the galaxy as defined by first order moments of the near-infrared light distribution (from the $F160W$ WFC image) is located at coordinates RA(J2000) = 14:53:07.86 and Dec(J2000) = -19:44:13.1, where we estimate the uncertainty of the galaxy centroid to be approximately 0\farc{15} in each coordinate. This position is 1\farc{3} distant from the radio position of the GRB RA(J2000) = 14:53:07.8078, Dec(J2000) = -19:44:11.995 as given in \citet{2016arXiv161006928M}.

\subsection{The stellar population at the GRB site}

\begin{figure}
\includegraphics[angle=0, width=0.99\columnwidth]{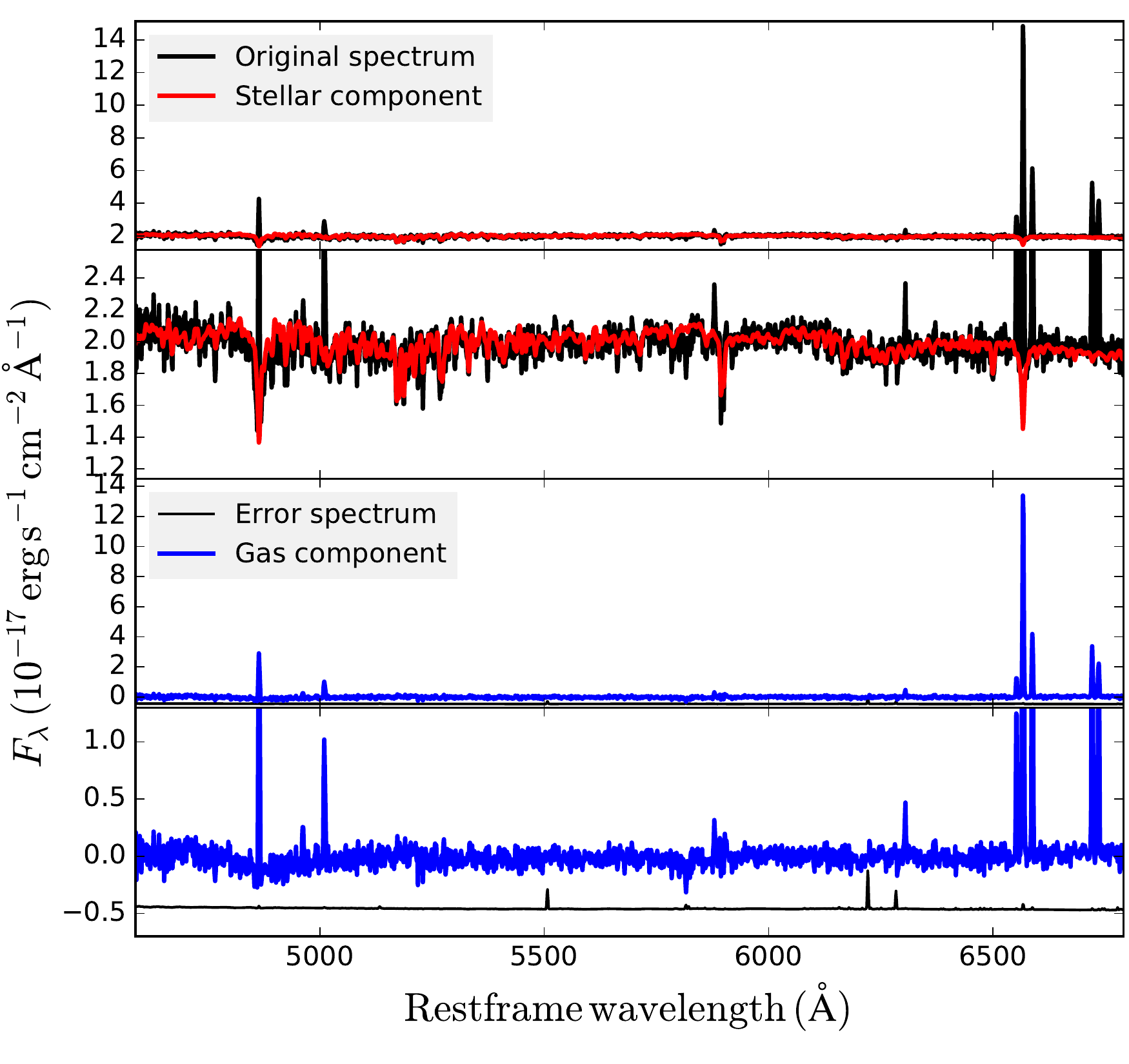}
\caption{Separating stellar and gas-phase components at the GRB explosion site. \textit{Top}: Original spectrum (black) extracted from the spaxel nearest to the GRB position and fitted stellar component (red). \textit{Second}: Zoom-in to the continuum. \textit{Third}: Resulting spectrum of the gas-phase contribution (blue). \textit{Bottom}: Zoom-in of the gas-phase contribution (blue) together with the error spectrum (black).}
\label{fig:starlight}
\end{figure}

As a first step in the analysis of the gas phase contribution, we model the stellar background to obtain reliable emission line fluxes in particular for the Balmer lines. In most regions of ESO 580-49 the stellar Balmer absorption is relatively strong, indicative of the presence of an evolved stellar population. Here, we follow closely our previous analysis of the galaxy hosting GRB~980425 \citep{2017A&A...602A..85K}, where we automatically fit the stellar component in individual spaxels using a combination of single stellar population models \citep{2003MNRAS.344.1000B} in \texttt{starlight} 
\citep{2005MNRAS.358..363C, 2009RMxAC..35..127C}. In our fits, we used a library of single stellar population synthesis models covering 15 stellar ages between 1~Myr to 13~Gyr, and three metallicities, 0.2~Z$_\odot$, 1 and 2.5~Z$_\odot$. From these stellar population models \texttt{starlight} produces a composite stellar continuum spectrum that best fits the data. The age of the dominant stellar populations are largely constrained by the strength of the stellar Balmer absorption, and the stellar population metallicities are constrained by the slope of the stellar continuum and the stellar absorption of metal lines. There is a degeneracy between stellar age and metallicity, such that old, metal poor stellar populations have similar spectra to young, metal-rich populations, which introduces a systematic uncertainty on these best-fit parameters on the order of $0.1-0.2$~dex \citep{2005MNRAS.358..363C}.

Fig.~\ref{fig:starlight} shows this procedure exemplified by a spectrum extracted at the GRB~111005A explosion site, where the Balmer lines can be clearly seen both in absorption and emission. The best-fit to the spectrum at the position of the GRB found with \texttt{starlight} corresponds to a model with a prominent stellar component (70\% of the observed the stellar light) with an estimated age of around 1 to 3~Gyr, as well as a smaller contribution (10-20\% of the stellar light) by a 10-20~Myr old stellar population. 

\subsection{The gas component}

\subsubsection{Ionization source}
\begin{figure}
\includegraphics[angle=0, width=0.99\columnwidth]{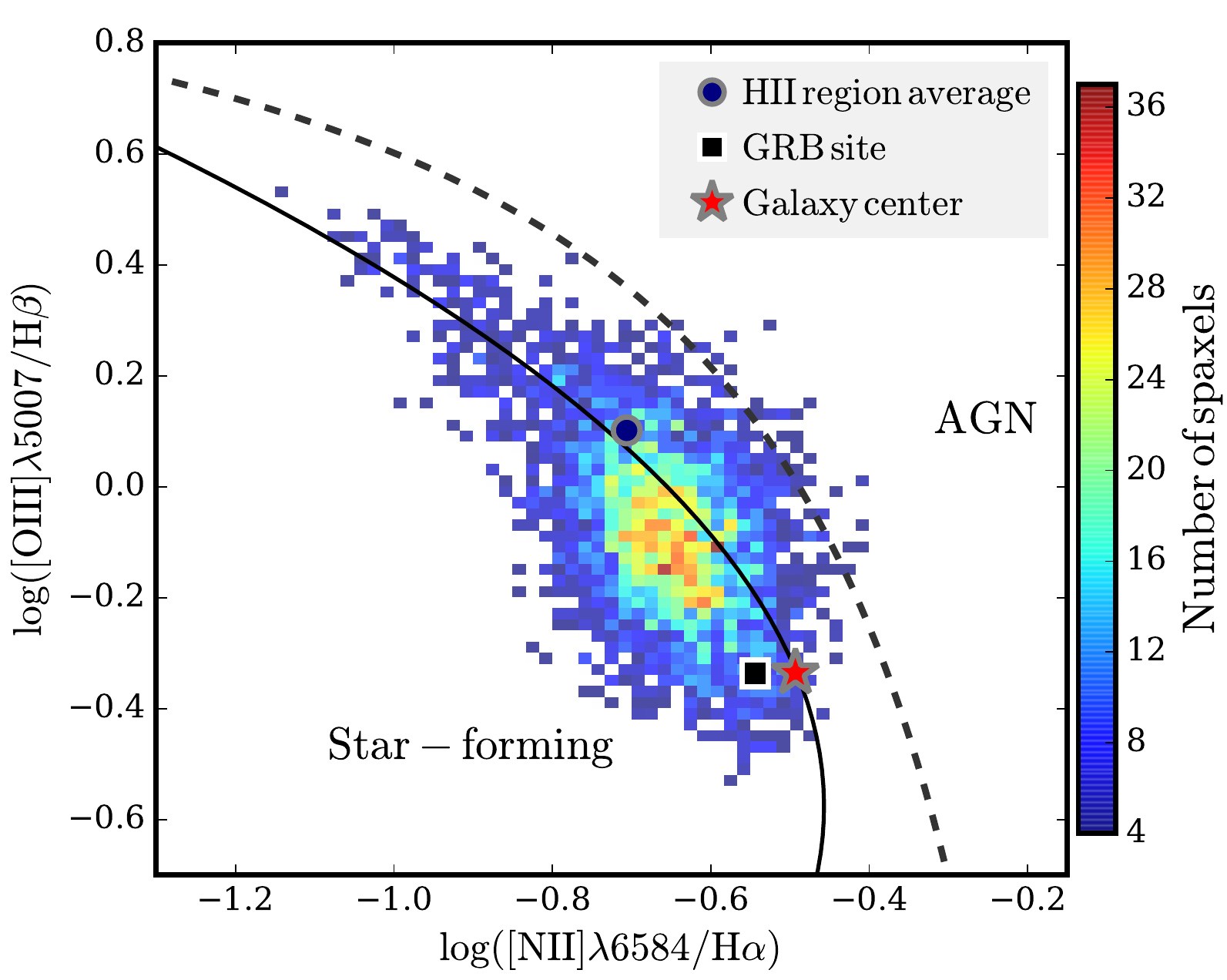}
\caption{Spaxel BPT diagram of ESO 580-49. Each data point corresponds to the spectrum of a single spaxel, where the individual emission lines are detected with a S/N of at least 3. The GRB explosion site, the galaxy centre, and the \hii\ region average are indicated by large symbols as shown in the legend. The differentiation line between AGN (or shock) and stellar ionization \citep{2013ApJ...774..100K} is shown by the dashed line. The solid line denotes the ridge line of local galaxies, i.e., the line with the highest density of SDSS galaxies \citep{2008MNRAS.385..769B}.}
\label{fig:bpt}
\end{figure}

A basic clue to the origin of the ionized gas is traditionally inferred through the BPT diagram named after \citet*{1981PASP...93....5B}. Here, two ratios of collisionally excited metal and Balmer lines (\oiii/\hb\ and \nii/\ha) allow us to discriminate the hard spectrum of an active galactic nucleus (AGN) from the black-body spectra of massive stars (Fig.~\ref{fig:bpt}).

All spaxels where we detect the respective lines with a  S/N of at least three are consistent with ionization from massive stars. The BPT diagram ascertains that no significant ionization from an AGN is present in ESO 580-49, even when considering the spaxels at the centre of the galaxy or at the transient position. Furthermore, the position of the spaxels suggest minimal contribution from diffuse ionized gas, which would otherwise lie in the bottom half of the `AGN' region of the diagram. The ionization properties at the GRB position and in the galaxy as a whole is thus fully consistent with origination from stellar excitation.

\subsubsection{Equivalent widths}
\label{sec:ew}

The equivalent width (EW) of emission lines such as \ha, \hb, or \hei\ is a rather direct proxy of the age $\tau$ in young starbursts at $\tau < 10$\ Myr \citep{1996ApJS..107..661S, 2003MNRAS.340...29C}. The specific relation between the two quantities EW and $\tau$ depends on the specific emission line and physical parameters like the star-formation history, the metallicity, initial-mass function, binarity and stellar rotation \citep[e.g.][]{1999ApJS..123....3L, 2009MNRAS.400.1019E}, but generally high EWs of $\mathrm{EW_{H\alpha}}\sim 100$~\AA\ or above are expected in the environments of long GRBs due to the short life time of their progenitors.

\begin{figure*}
\centering
\begin{subfigure}{.34\textwidth}
  \includegraphics[width=0.999\linewidth]{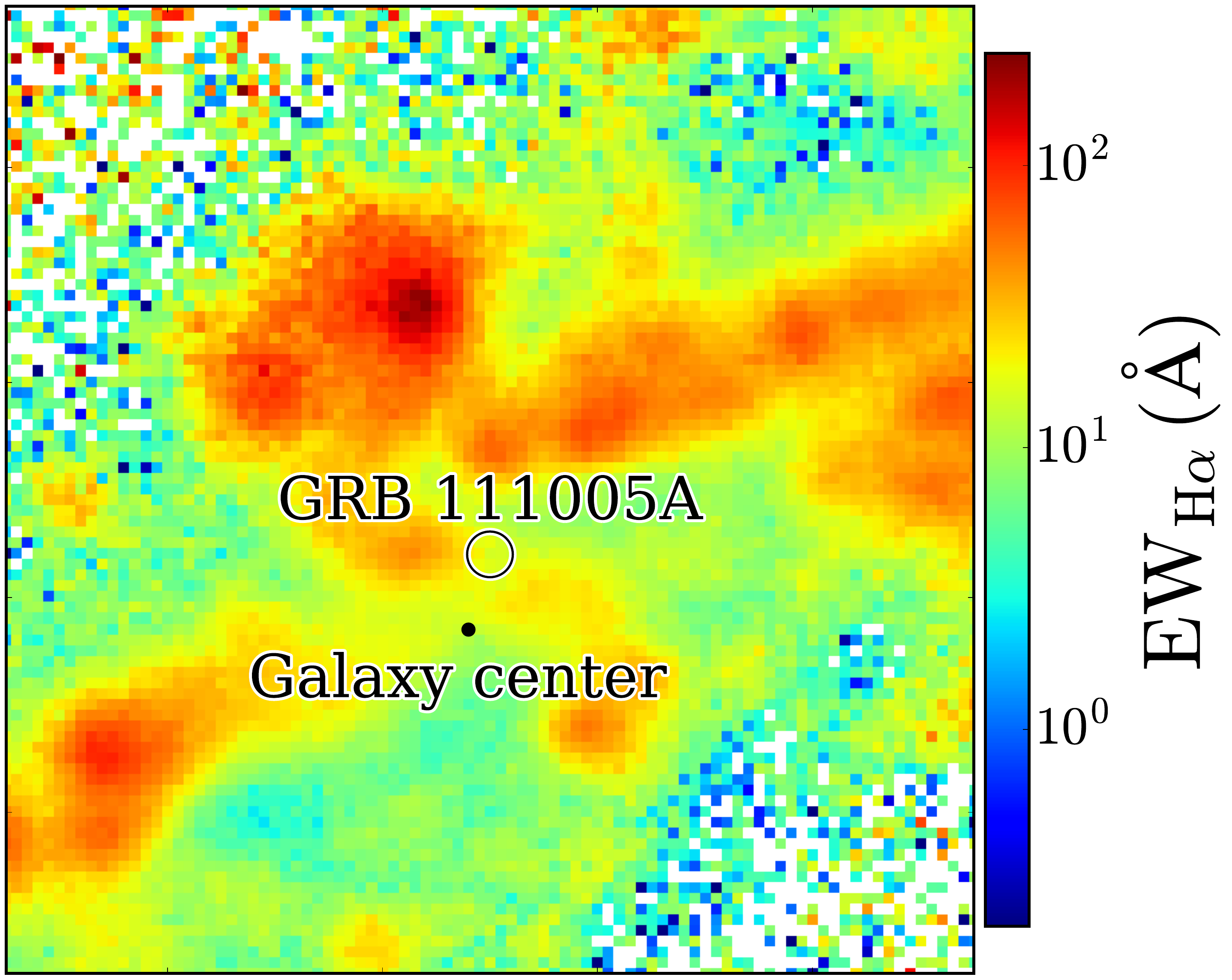}
\end{subfigure}
\begin{subfigure}{.34\textwidth}
  \includegraphics[width=0.999\linewidth]{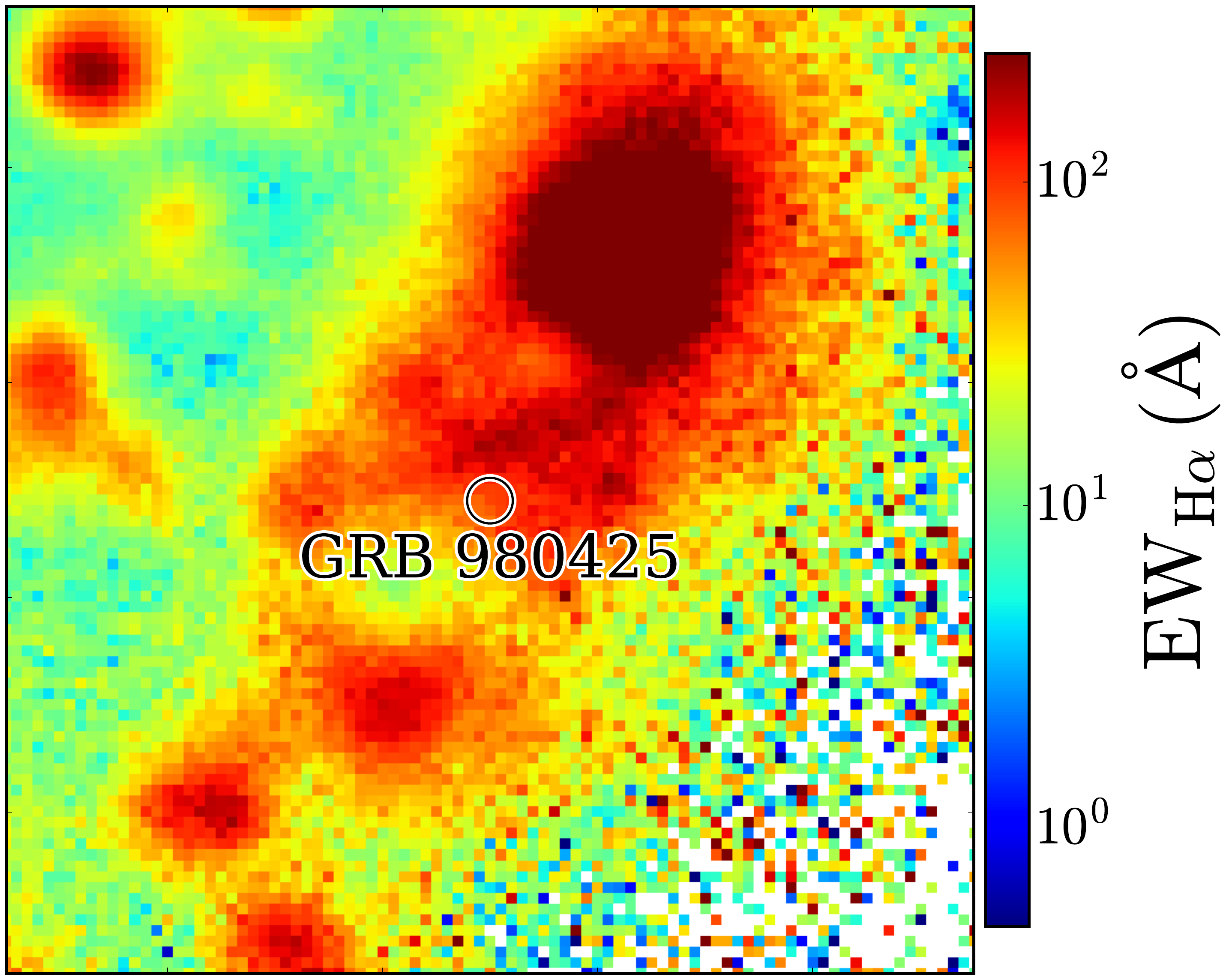}
\end{subfigure}
\begin{subfigure}{.34\textwidth}
  \includegraphics[width=0.999\linewidth]{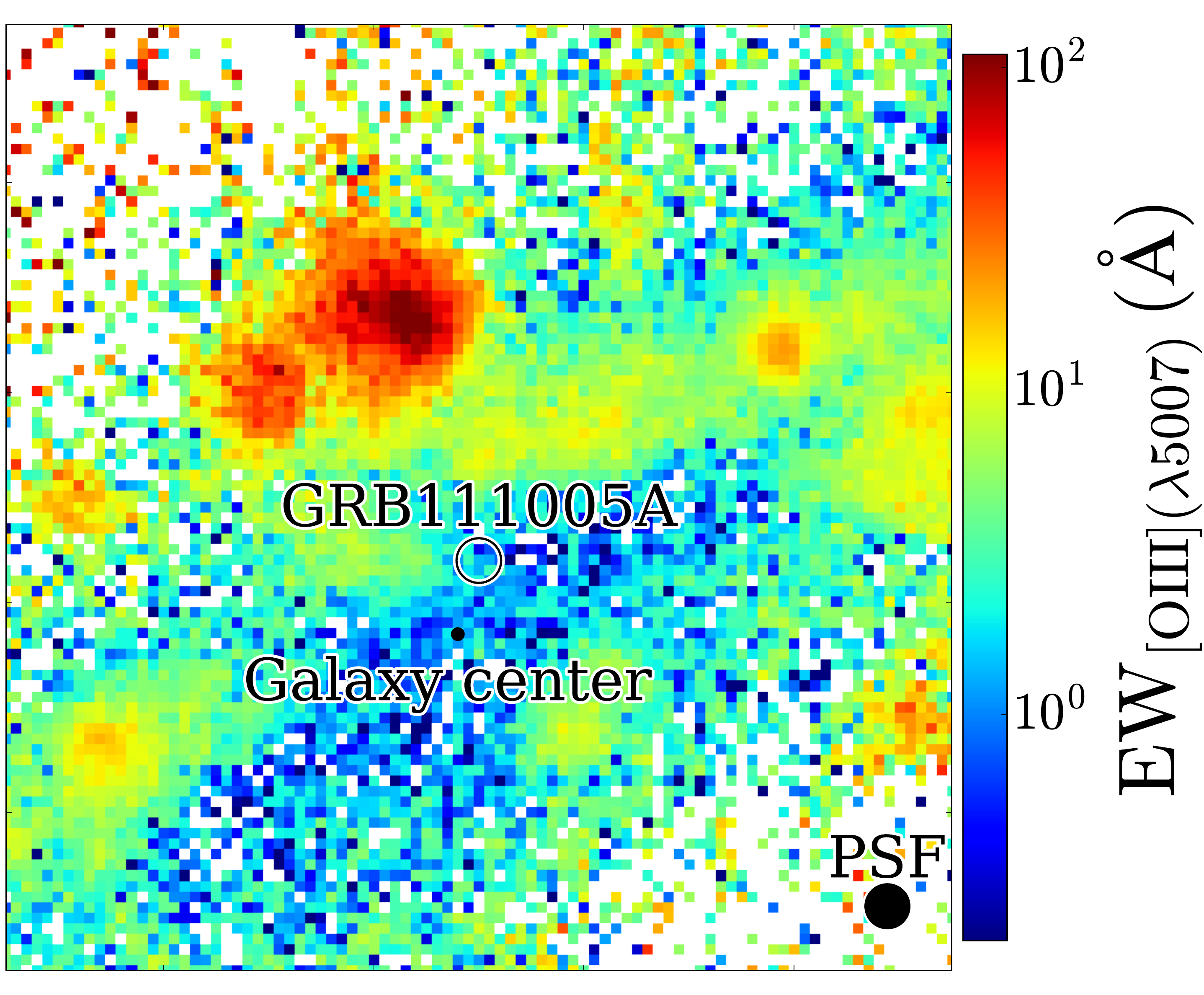}
\end{subfigure}
\begin{subfigure}{.34\textwidth}
  \includegraphics[width=0.999\linewidth]{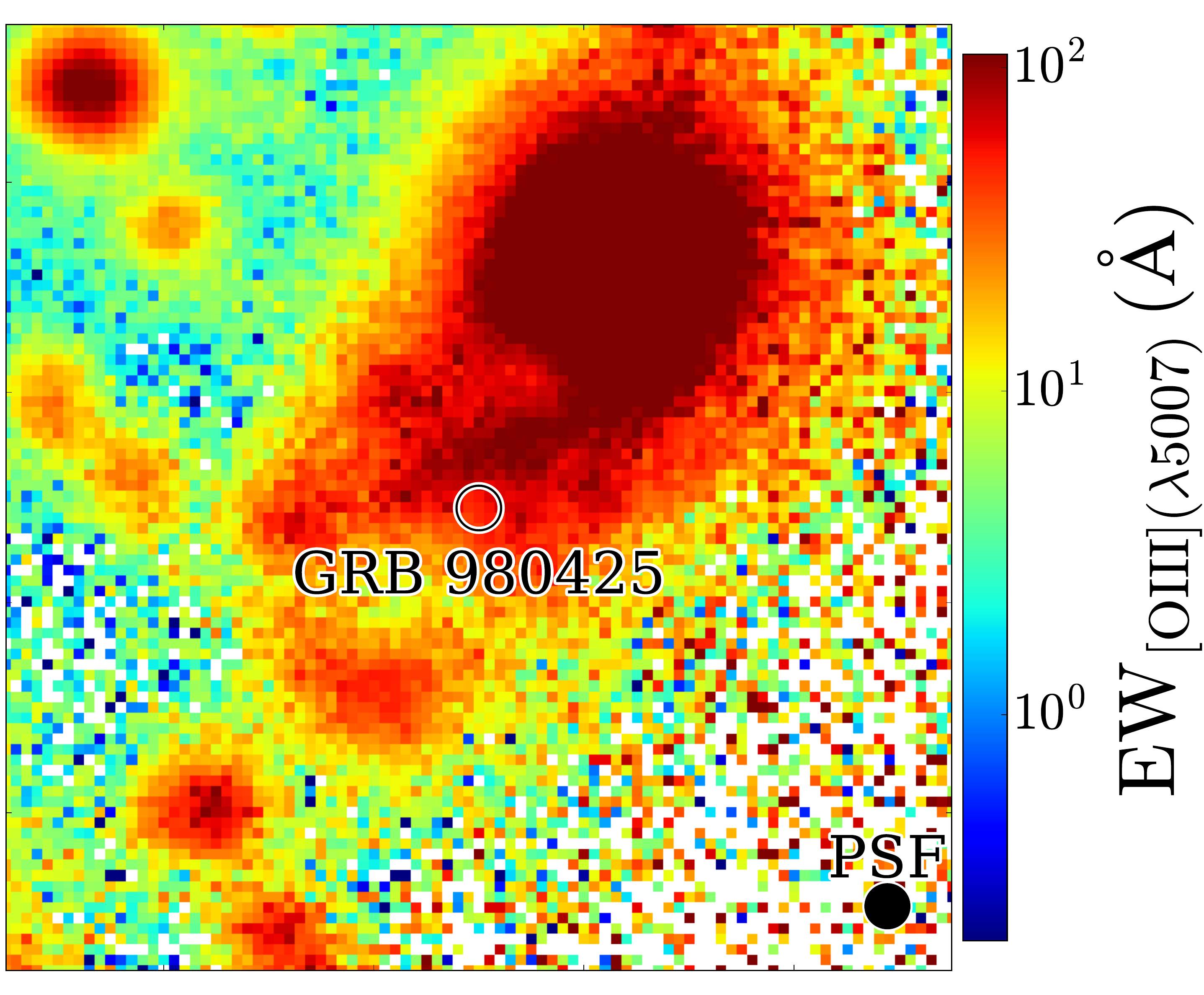}
\end{subfigure}

\caption{Comparison between the EW of \ha\ in the environment of the two, closest, long GRBs. GRB~111005A, at $D_L=62$~Mpc, is seen in the two \textit{left} panels, GRB~980425, at $D_L=39$~Mpc, in the two \textit{right} panels. The \textit{top} row shows \ha, the \textit{bottom} row \oiii($\lambda$5007). The galaxy centre of light is indicated by a black filled circle in the case of GRB~111005A. Each panel is approximately $20\times 20$~arcsec in size, corresponding to a physical scale of $4.7\times 4.7$~kpc and $3.5\times 3.5$~kpc for the fields of GRB~111005A and GRB~980425, respectively.}
\label{fig:ew}
\end{figure*}

Fig.~\ref{fig:ew} shows the equivalent width maps of two strong lines (\ha, \oiii($\lambda$5007)) typically detected in GRB host galaxies. As comparison to GRB~111005A we show similar maps from GRB~980425 with the same colour bar \citep{2017A&A...602A..85K}. The luminosity distance to GRB~980425 ($D_L=39$~Mpc) is comparable to the one of GRB~111005A ($D_L=62$~Mpc), but in contrast to SN-less GRB~111005A, GRB~980425 was associated with a luminous SN, the well-studied broad-lined Ic SN~1998bw \citep{1998Natur.395..670G, 2001ApJ...555..900P}.

Strong differences between both GRB environments exist, clearly apparent in the strength of both emission lines with respect to the continuum. While GRB~980425 was directly associated with a young star-forming region \citep{2000ApJ...542L..89F, 2017A&A...602A..85K}, such a link is absent in GRB~111005A. The EWs of \ha\ and \oiii\ at the GRB~111005A site are EW$_{\mathrm{H\alpha}} = 16\pm2$~\AA\ and EW$_{\mathrm{\oiii}}=2\pm1$~\AA, respectively, one to two orders of magniutde less than in the case of GRB~980425. Although the \ha\ EW is low, it is nevertheless larger than the threshold of around 3~\AA\ below which diffuse ionized gas is thought to dominate gas-phase emission \citep[e.g.][]{2011MNRAS.413.1687C, 2016MNRAS.461.3111B}. Therefore, despite the comparatively weak \ha\ emission, it is still likely to originate from excitation from stars, albeit an old stellar population.

The closest \hii\ region is at an angular distance of $\sim1^{\prime\prime}$ or a projected distance of 300~pc to the east, which is much larger than the size of typical \hii\ regions of a few, to a few tens of parsec. The only \hii\ region that appears similar to the GRB~980425 explosion site lies $\sim3\farc{5}$ to the North-East. At a projected distance of 1~kpc, this region is clearly too far away to be of relevance for the GRB explosion. Runaway stars within the Milky Way can have velocities of up to 200~km~s$^{-1}$ \citep{2001A&A...365...49H}, and there are also some known examples of `hyper-velocity' stars with velocities of over 500~km~s$^{-1}$ \citep[e.g.][]{2008A&A...483L..21H}. However, such highly rapidly moving stars are rare, and it therefore seems unlikely that the GRB progenitor, a rare event in itself, had such a high radial velocity. \cite{2011MNRAS.414.3501E} used binary population synthesis models to predict the typical kick velocity that the progenitors of core-collapse SNe and GRBs would receive at the explosion of their binary companions, and they found the typical velocities were in the range $10-20$~km~s$^{-1}$. It would therefore take such a progenitor at least 15-30~Myr to reach the GRB explosion site if it originated in the \hii\ region 300~pc to the east. This is older than expected from most long GRB progenitor models. Nevertheless, we cannot rule out that the GRB progenitor originated from the \hii\ region located at a projected distance of 300~pc to the east of the GRB. Either way, this does not alter the fact that the environmental properties in the vicinity of GRB~111005A are unusual for a long GRB. The \ha\ EW measured at the GRB explosion site and within the closest \hii\ region imply a progenitor age of $\tau\gtrsim10$~Myr, or equivalently a maximum progenitor mass of around $M_{ZAMS}<15~M_\odot$ (see section~\ref{sec:host}), which is much older, and less massive, than has been inferred from spatially resolved observations of more typical GRB-SNe \citep[e.g.][]{2017MNRAS.472.4480I, 2017A&A...602A..85K}.

\subsubsection{Dust map}
\label{sec:dustmap}

An obvious explanation for the absence of a luminous optical signal from a SN would be substantial amounts of dust along the sight line towards the GRB. In light of the non-detection in both the near-infrared and with \textit{Spitzer} \citep{2016arXiv161006928M}, the required column density of dust would need to be large to absorb the emission of a potential SN below the limits set by the respective observations (section~\ref{sec:snlim}). 

\begin{figure}
\includegraphics[angle=0, width=0.99\columnwidth]{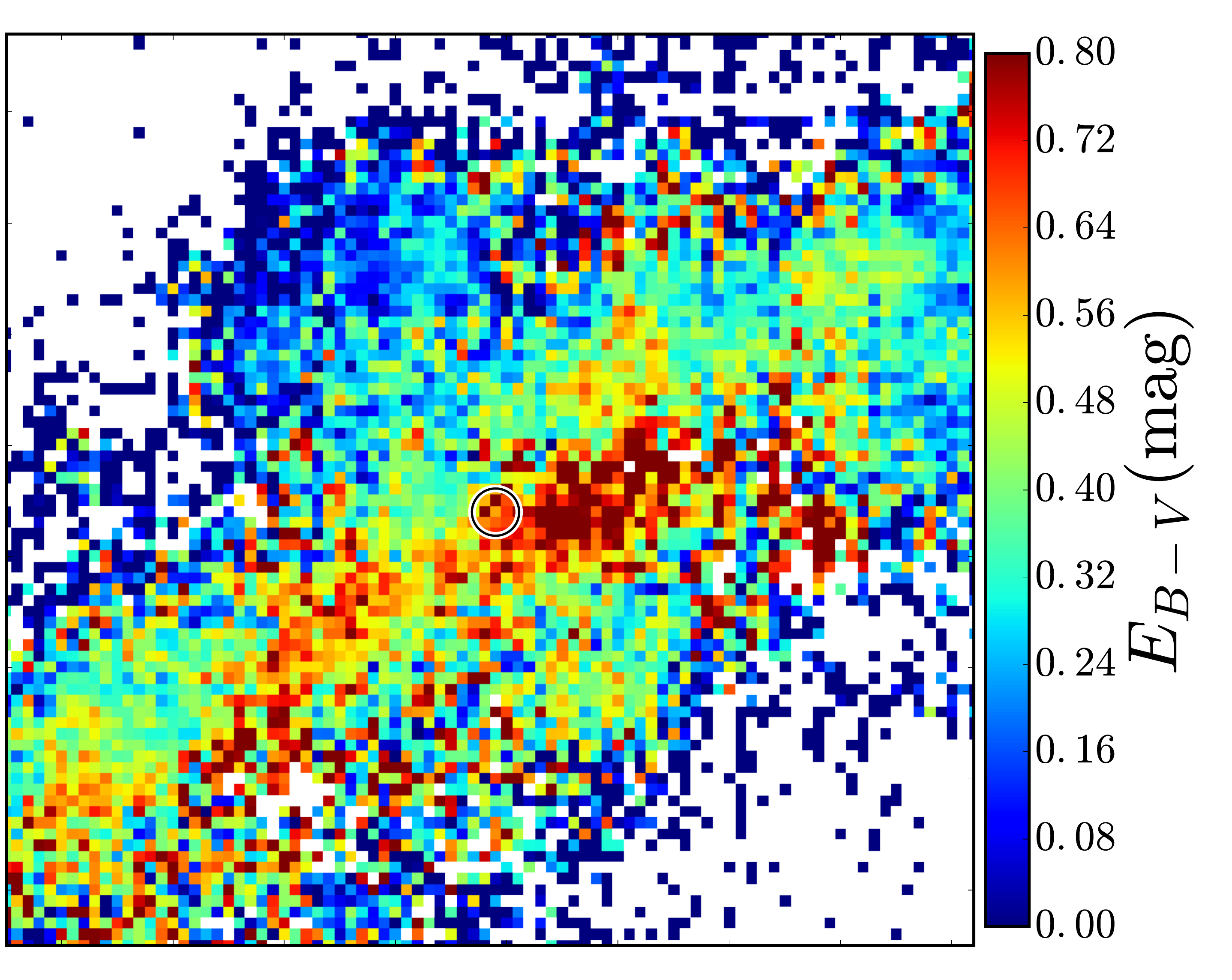}
\caption{Host galaxy dust reddening map, E(B-V), at the location of the radio transient associated with GRB 111005A. The position of the radio afterglow is marked with a black circle. The size of the region shown is $20\times 20$~arcsec, corresponding to a physical scale of $4.7\times 4.7$~kpc.}
\label{fig:av}
\end{figure}

To derive and measure the absorption by dust as parameterized by $A_V$ we follow a standard approach by measuring the difference between the observed \ha/\hb\ ratio to the theoretical value in case B recombination \citep{2006agna.book.....O}. Here, we assume a Milky-Way type reddening law, with an $R_V=3.08$, but note that this somewhat arbitrary choice has very little influence on our final result. Other extinction laws from the Local Group show very similar behaviour in the wavelength range of \ha\ and \hb, so that the estimates of $E_{B-V}$ on Fig.~\ref{fig:av} illustrate the typical visual absorption towards the ionized gas.

The reddening at the GRB position is significant, and among the largest observed values within the galaxy. The value derived from the spaxel closest the GRB site is $E_{B-V}=0.59\pm0.08$~mag or $A_V=1.8\pm0.3$~mag. Similar values are obtained from the environment: within a radius of 170~pc of the GRB position, we measure $E_{B-V}=0.66\pm0.12$~mag or $A_V=2.0\pm0.4$~mag.

While these measurements are not sufficiently large to explain the absence of the SN due to absorption by dust, they are in stark contrast to measurements obtained in a similar manner for the GRB~980425 environment ($A_V\sim0.1$~mag). It is important to note that the galaxy is oriented edge-on, and thus one could speculate about the possibility that the GRB occurred at the far side of the galaxy, and that a large cumulative dust column thus fully extinguishes all stellar light associated with the GRB natal \hii\ region, eluding its detection in our MUSE maps. For the dust depleted SN curves in Fig.~\ref{fig:sn} to lie below our detected afterglow upper limits in all bands, the integrated dust column would have to have a reddening of $E_{B-V}>10$.
 
Our data allows the detection of \hb\ emission at the position of the GRB to an $E_{B-V}$ of up to $\sim$2, above which the \hb\ flux would be detected at a S/N of less than three. We measure no regions in the galaxy with $E_{B-V} > 1.0$ nor do we observe a sharp cut-off in the host galaxy $E_{B-V}$ distribution, which instead peaks at $E_{B-V}\sim 0.2$, and then decreases steadily at higher $E_{B-V}$ values. Furthermore, even in the absence of \hb, we should be able to detect \ha\ emission from any young \hii\ region that is co-spatial with the GRB position with a dust reddening of up to $E_{B-V}\sim 3$. We therefore consider it unlikely that there exists a very heavily extinguished region ($E_{B-V} > 10.0$) at the location of the GRB.

\subsubsection{Metallicity}
\label{sec:metalmap}

The final parameter that we consider is metallicity, traditionally expressed in the oxygen abundance $\oh$, where the solar value is $\oh=8.69$ \citep{2009ARA&A..47..481A}. Metallicity has theoretically been a key parameter in GRB formation: in single progenitor models, a minimum metallicity is required for the hydrogen and helium envelope to be stripped, yet a too high value (above $0.3~Z_{\odot}$) would also remove too much angular momentum for an accretion disk to form. 

GRB hosts in general are characterized by a distribution of $\oh$ peaking at around $\oh= 8.4$ or $0.5\,Z_{\odot}$ at $z < 1$, which can be explained as the result of GRBs following star-formation, while at the same time avoiding the most metal-rich galaxies \citep{2013ApJ...774..119G, 2015A&A...581A.125K}. Metal-enriched GRB hosts with oxygen abundances close to or at the solar value do exist \citep{2012A&A...546A...8K, 2013A&A...556A..23E, 2015A&A...579A.126S}, but they are much less frequent than one would expect based on the contribution of similar types of galaxies to the global star-formation rate \citep{2015A&A...581A.102V, 2016ApJ...817....8P}. It thus seems that qualitatively, the theoretical expectation that a high metallicity prevents GRB formation is consistent with observations, while quantitatively, there appears to be some mismatch between the actually observed and theoretically expected values. 

It is clear, however, that the measurements of oxygen abundance in GRB hosts might not directly trace the progenitor metallicity. Most of the values present in the literature are derived from spectra that are integrated over the whole galaxy. In the few cases where spatially resolved spectral data have been acquired, either from an appropriately placed slit \citep[e.g.][]{2008ApJ...676.1151T,2015A&A...579A.126S}, or from IFU observations \cite[e.g.][]{2014MNRAS.441.2034T,2017MNRAS.472.4480I,2017A&A...602A..85K}, the metallicity has typically been measured from spatial regions $>500$~pc. The only exception to this was in the case of the closest GRB known to date, GRB~980425, where \citet{2017A&A...602A..85K} used MUSE data to study the nebular emission line properties of the host galaxy on scales of 100~pc. Also, which of the various emission line diagnostics available are most appropriate to infer $\oh$ is a matter of constant debate in the literature \citep[e.g.][]{2017ApJ...849L...4C,2018A&A...610A..11I}, and it is not clear which (if any) of the many proposed strong-line ratios yields accurate metallicities on absolute scales \citep{2008ApJ...681.1183K}.

\begin{figure}
\includegraphics[angle=0, width=0.99\columnwidth]{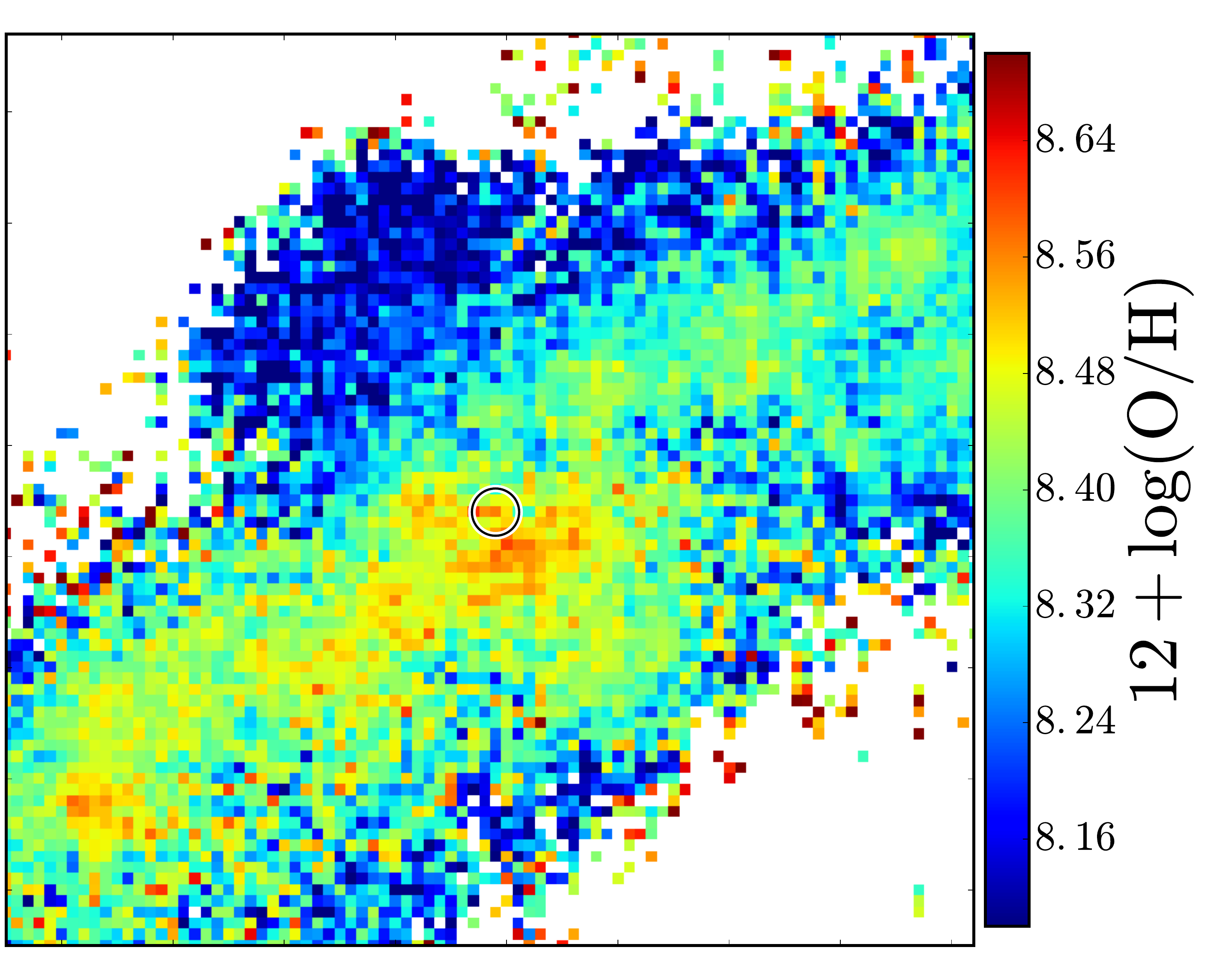}
\caption{Host galaxy oxygen abundance map at the location of the radio transient associated with GRB 111005A. The metallicity was derived using the diagnostic from \cite{2016Ap&SS.361...61D}. The position of the radio afterglow is marked with a black circle. The size of the region shown is the same as for Figs.~\ref{fig:ew} and \ref{fig:av}.}
\label{fig:oh}
\end{figure}
In any case, GRB~111005A provides a rare example where we can obtain a spatially resolved oxygen abundance measurement from the ionized gas at the GRB explosion site. Given the short life time of possible GRB progenitors, this is as close as currently possible to the actual GRB progenitor metallicity. Fig.~\ref{fig:oh} contains a map of the oxygen abundance of the GRB environment in the diagnostic ratio of \citet{2016Ap&SS.361...61D}, and we measure $\oh=8.56\pm0.04$ at the GRB site. Different diagnostics return generally compatible values, in particular when the significant systematic errors of these diagnostics are taken into account. For example, the so-called 'O3N2' metallicity diagnostic \citep{2004MNRAS.348L..59P} gives a value of $\oh=8.63\pm0.03$ at the explosion site. We conclude that the GRB environment is substantially enriched with metals, and has a metallicity of around or possibly somewhat lower than the solar value.

\subsubsection{Velocity Map}
\label{sec:velmap}

\begin{figure}
\includegraphics[angle=0, width=0.99\columnwidth]{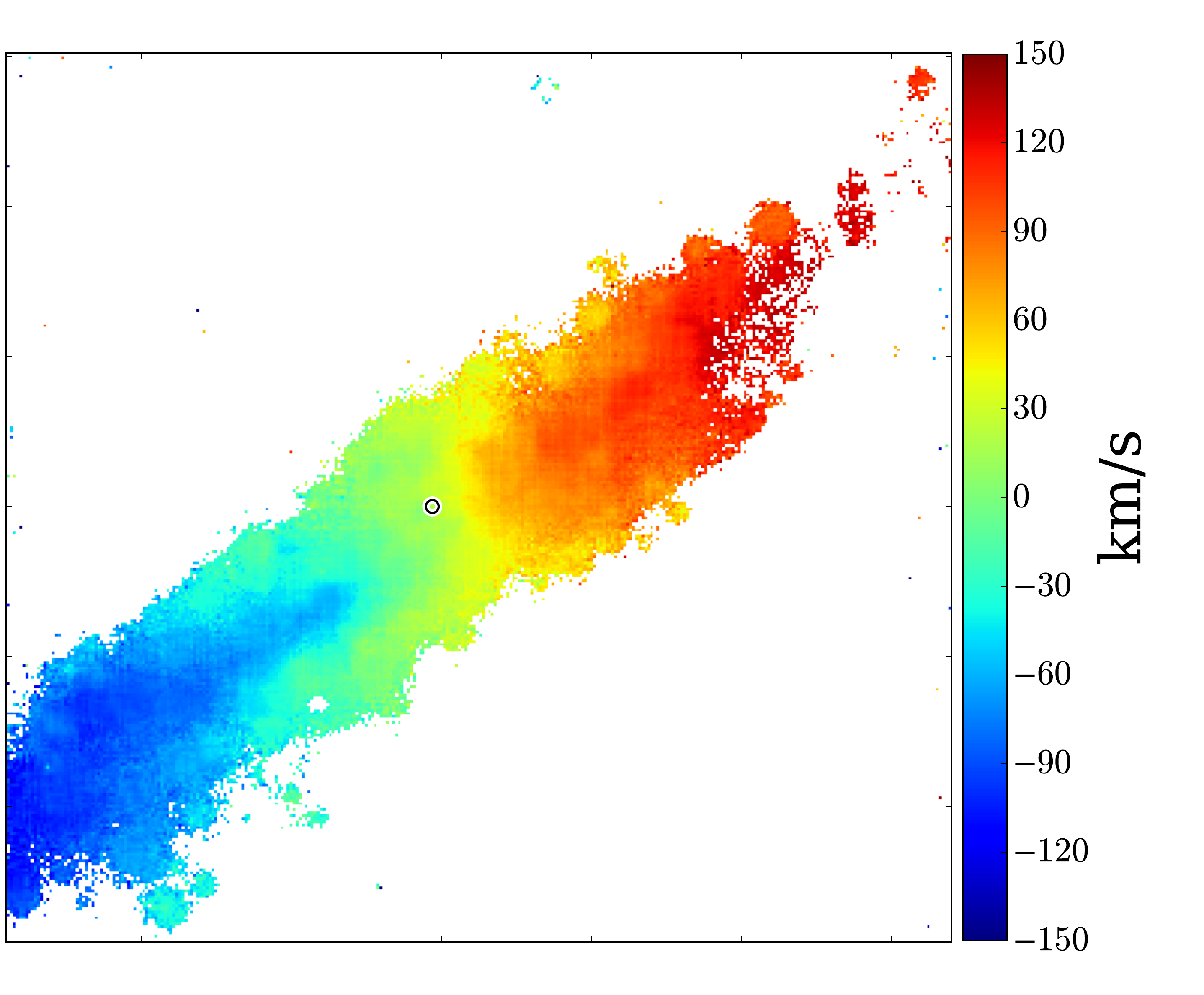}
\caption{Velocity map inferred from the position of the \ha\ line relative to a redshift of $z=0.01326$. The position of the GRB is indicated by the small circle in the middle of the figure.}
\label{fig:vel}
\end{figure}

In order to investigate the possibility that the GRB arose in a foreground satellite galaxy, or from a lower metallicity galaxy that may have merged with ESO 580-49, we studied the galaxy velocity distribution of ESO 580-49. In Fig.~\ref{fig:vel} we show the velocity map, which shows the velocity offset of the \ha\ line relative to a redshift of $z=0.01326$. The velocity map shows a clear rotating disk, with little indication for any kinematic disturbance at either the GRB position, or in any other parts of the galaxy.

\section{Summary and conclusions}
\label{sec:dis}

\subsection{GRB~111005A in context}

Our analysis of the \emph{Swift}-BAT data shows that GRB~111005A is a long-soft and low-luminosity GRB. A radio transient that is likely the radio afterglow of GRB~111005A was detected a few days after the GRB trigger \citep{2016arXiv161006928M}, and it was monitored for approximately 100 days after the \emph{Swift}/BAT trigger. The radio afterglow was close to (but not at) the centre of the galaxy ESO 580-49 at $z = 0.0133$, and using conservative estimates on the probability that the GRB, the radio transient and the low-redshift galaxy are unrelated, we find that their association is credible. The lack of any evidence for AGN emission from the position of the galaxy on the BPT diagram (Fig.~\ref{fig:bpt}) rules out that an AGN produced the radio emission, further strengthening the association between the radio transient and GRB~111005A. This makes GRB~111005A the second closest long GRB ever detected.

Long duration GRBs are expected to be followed by a supernova but observations with GROND, VLBA and \textit{Spitzer} confirm the lack of a luminous supernova associated with this GRB. Even when accounting for a significant amount of dust along the sightline (our best estimate is $A_V\sim2$~mag from Section~\ref{sec:dustmap}), we find that any SN associated to GRB~111005A is fainter than SN1998bw by at least 3 magnitudes in the $r$-band and 4 magnitudes in the $K$-band. These deep limits imply that the $^{56}Ni$ mass synthesized in any SN explosion must be less than $10^{-3}$ $M_{\odot}$, and thus several orders of magnitudes lower than in regular GRB/SNe \citep[e.g.][]{2017MNRAS.469.2498M}.

We thus conclude, in a similar way to \citet{2016arXiv161006928M}, that GRB~111005A is a low-$z$, SN-less GRB, similar to GRB~060505 ($z = 0.089$) and GRB~060614 ($z = 0.125$) \citep[e.g.][]{2006Natur.444.1047}, but at a much closer distance, and with an unambiguous long GRB classification from the prompt emission. 

The question that immediately arises is thus: Which progenitor scenario is able to account for a temporally-long GRB, but which lacks the supernova that is expected to arise simultaneously? It has been suggested that GRB~060505 and GRB~060614 could result from the direct collapse of a massive star to a black hole \citep[e.g.][]{2006Natur.444.1047}, or from a central engine that lives long enough to feed the GRB, but that does not involve a massive star \citep[e.g.][]{2006Natur.444.1053G}. A further window into understanding the nature of these GRBs is provided by studying their host galaxies \citep[e.g.][]{2008ApJ...676.1151T, 2014MNRAS.441.2034T}. In contrast to the two examples from 2006, GRB~111005A is close enough to resolve \hii-regions from the ground with modern integral-field spectrographs, and we can thus study here its environment in exquisite detail.\footnote{\citet{Niino2015MNRAS} argue that a FWHM of less than $500\times500$~pc$^{2}$ is required to measure a metallicity that is representative of the transient environment. In the case of GRB~060505 and GRB~060614, equivalent MUSE data would provide a resolution larger than 1~kpc$^{2}$.}

\subsection{The host galaxy in context}
\label{sec:host}
The global properties of long GRB host galaxies have shown them to be young, star-forming galaxies deficient in metals, strengthening the evidence that GRBs are associated with metal-poor massive stars, which is apparently necessary to produce a GRB from a core-collapse event. In contrast to this, we find that the host of GRB~111005A is a moderately star-forming, relatively metal-rich and dusty galaxy, and thus at odds with the general population of GRB hosts and the requirement of GRB progenitors. Our MUSE-IFU observations provide one of the best-resolved studies of a GRB host and the GRB environment so far, and although we are limited in our resolution along the radial-axis due to the edge-on orientation of the galaxy, it is still possible to resolve individual \hii-regions. From these data we thus find that the GRB is not located within a region of ongoing star-formation, and its immediate environment has a near solar metallicity within the nearest \hii-region to the radio afterglow (Figs.~\ref{fig:ew} and ~\ref{fig:oh}). Although the absolute metallicity at the location of the GRB is somewhat uncertain due to the edge-on orientation of the galaxy, the lack of ongoing star formation is a robust result. The likelihood, that there is a region of star formation at the GRB position that is unseen due to dust extinction is small given the sensitivity of our data and the dust distribution that we measure (see Fig.~\ref{fig:av} and section~\ref{sec:dustmap}).

From nebular line emission maps produced from the MUSE datacube, we find that there is no \hii-region within 300~pc of the transient position, and even this `nearby' \hii-region is far less star-forming than the \hii-region associated with GRB~980425 (Fig.~\ref{fig:ew}). It is also in stark contrast with other nearby, long GRBs, that are almost exclusively found in the UV-brightest region of their host galaxies \citep{2006Natur.441..463F, 2016ApJ...817..144B, 2017MNRAS.467.1795L}, as expected if long GRBs are produced from the core collapse of a massive star. If this GRB had been at a redshift $z=1-2$, its host galaxy would be unresolved. To check how the physical properties that we would measure would change in this case, we derived the galaxy integrated metallicity over the whole MUSE field of view. This gave us a metallicity of $\oh\sim 8.4$ when using the \cite{2016Ap&SS.361...61D} diagnostic, and an O3N2 metallicity of $\oh\sim 8.6$. This is very similar to the metallicities measured at the GRB site (see section~\ref{sec:metalmap}), and it indicates that if the GRB had been at $z=1-2$ we would still measure its host galaxy environment to be metal-rich.

At the location of the radio afterglow there is no detectable emission from \oiii, as would be expected from a very young stellar population. The \ha\ emission at the GRB position is also relatively weak, and the measured small equivalent width EW$_{\mathrm{H\alpha}}=16\pm 2 $ \AA\ implies an underlying stellar population that is older than approximately 10~Myr \citep[and references therein]{2016arXiv160703446K}. The age limit for the stellar population then implies a maximum progenitor mass of around $M_{\mathrm{ZAMS}}<15~M_\odot$ \citep{2000A&A...361..101M}. The combination of solar metallicity and small \ha\ equivalent width at the GRB position are in firm contradiction with the requirements of conventional single-star core-collapse GRB progenitor models. The properties of the GRB nearby environment and the mass constraints on the progenitor are similar to those of collapsar supernovae of Type II-p \citep{2003ApJ...591..288H}, although SNe type II-p have typical $R$-band peak magnitudes between $M_R=-16$ mag and $M_R=-18$ mag \citep{2014MNRAS.442..844F}, which is $3-5$ magnitudes brighter than our photometric upper limits (see section~\ref{sec:snlim}). Nevertheless, there are examples of low-luminosity SNe type II-p that have absolute magnitudes $M_R\sim -14$ mag \citep[e.g.][]{2004MNRAS.347...74P}, and there was one exceptional transient event with $M_R=-12$ mag \citep{2007Natur.447..458K} that has been proposed to be a SN type II-p \citep{2007Natur.449E...1P}. Nevertheless, unlike GRB-SNe, the progenitors of Type II-p keep their hydrogen envelope, which is evident in the detection of strong \ha\ emission lines in their spectra. How the GRB jet would be able to pierce through the stellar hydrogen envelope is still to be demonstrated theoretically.

\subsection{The progenitor properties of GRB~111005A}
The host galaxy observations of GRB~111005A presented here provide very stringent constraints on the properties of the progenitor that gave rise to GRB~111005A, which effectively rule out the most popular single-star collapsar model \citep{Woosley2006ApJ}. Any proposed progenitor system has to be relatively insensitive to environmental metallicity, it must produce at most a very dim SN, and most importantly of all, it cannot be associated with a very young stellar population. 

Metallicity independent channels with binary systems have been discussed for the formation of GRBs ~\citep{Podsiadlowski2010MNRAS, WoosleyHeger2012ApJ, deMink2009AA, Song2016AA}. \citet{Podsiadlowski2010MNRAS} demonstrated that a helium star in a binary system can be tidally locked with a low-mass star, which could lead to transfer of mass from the low-mass star to the high-mass helium star. This would then enable the expulsion of the outer hydrogen- and helium-rich common envelope, leading to the formation of a rapidly spinning carbon core. For a sufficiently large angular momentum the core could collapse to form a GRB. Similarly, \citet{deMink2009AA} and \citet{Song2016AA} show that massive stars in overcontact binary systems may interact tidally to evolve into chemically homogeneous systems. The end phases are likely to be C- or N-rich Wolf-Rayet stars and thus candidates for GRB progenitors.
In such a scenario metallicity does not play a dominant role, but still a massive primary ($M\sim18~M_{\odot}$) is required for the GRB explosion. This is at odds with the absence of a SN and the lack of massive stars in the vicinity of GRB~111005A, but clearly most of the theoretical scenarios invoked to date to explain the GRB phenomenon are designed to explain the GRB/SN connection, and not its absence.

Similarly, progenitor models that invoke the direct collapse of a massive star to a black hole to explain SN-less GRBs seem unfeasible in this case. In particular the absence of a luminous \hii~region associated with GRB~111005A, and as a consequence the relatively mature population of stars at the GRB position, argues against such a scenario.

A more applicable scenario seems here the suggestions of \citet{2006Natur.444.1044G} and \citet{2006Natur.444.1053G}, proposing a genuinely novel explosion mechanism for GRB~060614. This might possibly involve compact object mergers rather than common envelope/over contact binaries, which is usually the preferred progenitor scenario for short GRBs \citep{Narayan1992ApJ,Mochkovitch1993Nature,2014ARA&A..52...43B,2017ApJ...848L..12A}. In the case of GRB~060614, further evidence for a novel explosion mechanism lise in the claimed detection of a macronova - a late time optical/near-infrared re-brightening due to the decay of radioactive elements of the Lanthanide series. This is a process that is unique to merger events \citep{2015NatCo...6E7323Y, 2015ApJ...811L..22J}. Although compact objects such as neutron star - neutron star mergers and neutron star - black hole mergers can produce relativistic jets for short periods of time, and the prompt emission and environmental properties of GRB~111005A presented here may motivate the development of similar models but where the central engine can remain active for longer than a few seconds.   

Our very detailed MUSE host galaxy data thus imply that a new progenitor model is required to account for the properties observed in GRB~111005A, which combined with GRB 060505 and GRB 060614, may thus form a new class of SN-less long duration-soft GRBs. All previously suggested single-star as well as common-envelope and over-contact binary evolution in relatively massive stars ($\ge$ 18 M$_{\odot}$) have been suggested in order to provide metallicity-independent progenitor evolution channels and to explain the GRB-SN association rather than the lack of it. Apart from these scenarios, further exploration of novel mechanisms for evolution of compact binaries could provide further clues to metallicity independent and SN-less long duration GRB channels.  

\begin{acknowledgements}

We thank the referee for their constructive comments and their sharp eye. P.S., J.F.G., T.K., and M.T. acknowledge support through the Sofja Kovalevskaja Award to P. Schady from the Alexander von Humboldt Foundation of Germany. DAK acknowledges financial support from the Spanish research project AYA 2014-58381-P, and from Juan de la Cierva Incorporaci\'on fellowship IJCI-2015-26153. Part of the funding for GROND (both hardware as well as personnel) was  generously granted from the Leibniz-Prize to Prof. G. Hasinger (DFG grant HA 1850/28-1).

\end{acknowledgements}

\bibliography{./bibtex/refs}

\begin{appendix}

\end{appendix}
\end{document}